\documentstyle[psfig19,floats]{lamuphys}
\makeatletter
\let\chapter\hid@chapter
\makeatother
\def\begineq{\begin{equation}}
\def\endeq{\end{equation}}

\begin{document}
\pagenumbering{arabic}
\title{Why air bubbles in water glow so easily}
 
\author{Michael\,P.\,Brenner\inst{1}, 
Sascha\,Hilgenfeldt\inst{2},
and Detlef\,Lohse\inst{2}}
 
\institute
{Department of Mathematics, \\ Massachusetts 
Institute of Technology, \\ Cambridge, MA 02139, USA \\
\and
Fachbereich Physik, Philipps-Universit\"at Marburg, \\ Renthof 6,
D-35032 Marburg, Germany}
 
\maketitle
 
\begin{abstract}
Sound driven gas bubbles in water can emit light pulses
(sonoluminescence). 
Experiments show a strong dependence on the type of gas 
dissolved in water.
Air is found to be one of the most friendly gases 
towards this
phenomenon. Recently,
\cite{loh96} have suggested 
a chemical mechanism to account for the strong 
dependence
on the gas mixture: the dissociation of nitrogen
at high temperatures and
its subsequent 
chemical reactions to highly water soluble gases
such as NO, NO$_2$, and/or NH$_3$.
Here, we analyze the consequences 
of the theory and offer detailed comparison with
the experimental data of Putterman's UCLA group.
We can quantitatively account for
heretofore unexplained results.
In particular, we understand why the argon percentage
in air is so essential for the observation of stable SL.
\end{abstract}


\section{Introduction}

\cite{gai90}
discovered that a gas bubble levitated in water by
a strong periodically modulated acoustic field can emit 
bursts of
visible light so intense as to be observable to
the naked eye, a phenomenon called
single bubble sonoluminescence (SL).
Previous workers had observed sonoluminescence
in multibubble cavitation clouds:  when a liquid is 
subjected to
large negative pressures, the liquid rips apart into a 
cloud of bubbles,
which then collapse violently.
 \cite{fre34}
placed a photographic plate in the vicinity 
of a cavitating bubble cloud, and discovered that it was 
exposed to
a low level of radiation.
Single bubble SL is distinguished 
from multibubble
sonoluminescene not only by the much higher
light intensity,
but also by the fact that the bubble remains in the 
oscillating state
for billions of oscillation periods (days) without 
dissolving or changing
its average size.

Soon after Gaitan's initial discovery,
Putterman's research group at UCLA began
to uncover a myriad of interesting dynamical properties 
of single bubble SL
(\cite{bar91,hil92,hil94,hil95,bar94,bar95,loe93,loe95,wen95}).
The width
of the light pulse is less than $50$ picoseconds.  The 
accuracy of the
flashes differs by less than $50$ picoseconds from cycle 
to cycle.  The
phenomenon shows strong dependence on the 
experimental parameters
as the forcing pressure amplitude $P_{\rm a}$, the mass concentration 
$c_\infty$ of dissolved gas in
the liquid far away from the bubble, and the temperature of the 
liquid.
SL is only observed for  $P_{\rm a} \sim 1.1$\,atm -- $1.5$\,atm (for an ambient
pressure of $P_0=1$\,atm) for all gases. The dependence on 
$c_\infty$ is more subtle. For pure argon gas {\it stable} SL is observed
only in a small window of 
extremely low concentration around
$c^{\rm Ar}_\infty/c_0 \sim 0.4\%$, whereas
for air the water has to be degassed only down to $c^{\rm air}_\infty/c_0 \sim 10
- 20\%$ to obtain stable SL.
Here, $c_0$ is the saturation 
concentration of the gas in the liquid. 
Unstable SL, characterized by a
``dancing bubble'' and by oscillations in the phase and intensity of the
light pulse, is observed in both cases (air and argon) for larger
$c_\infty/c_0$. For pure
nitrogen bubbles no stable SL is observed at all, only
very weak unstable SL. 
Recently,
\cite{wen95}
also
observed single bubble SL in nonaqueous fluids (alcohols) where the
phenomenon shows 
a strong dependence
on the type of liquid.

The major challenge in understanding the SL 
experiments is to determine
how much of the phenomenology is hydrodynamic,
how much is connected with the atomic properties of the 
gases,
and how much stems
from chemistry.
To allow for detailed comparison between experiment and
theory,
 \cite{hil96}
 calculated {\it phase diagrams}
which depict domains of different bubble behaviors.
They must depend
on parameters which are easily experimentally adjustable.
These are the forcing pressure $P_{\rm a}$ and the gas concentration
$c_\infty$.
As emphasized first by 
\cite{hil94} the ambient radius
$R_0$ (i.e., the radius of the bubble under standard ambient conditions;
$R_0\sim 5\mu{\rm m}$ in the SL experiments)
is not an experimentally controllable parameter 
but adjusts itself dynamically.

We will see that much of above phenomenology can already be understood
from the hydrodynamics and the stability of the bubble. Three types of
instabilities will turn out to be important: (i) Shape instabilities of the
bubble (\cite{ple54,str71,pro77,ple77,bre95,hil96}),
(ii) diffusional instabilities
(\cite{cru94b,loe93,bre96,hil96}), and
(iii) chemical instabilities (\cite{loh96}).
\cite{hil96}  consider   only the instabilities
(i) and (ii). They could
quantitatively understand the experimental
results (\cite{bar94,bar95,loe95})
 for pure argon bubbles, but not
for gas mixtures with {\it molecular} constituents as e.g.\ air.
For air the theoretical
phase diagrams look very similar to those of argon,
but experimentally (\cite{bar94,bar95,loe93}) stable SL is found for
about 100 times larger gas concentration, as already mentioned above.
In this paper we elaborate our idea (\cite{loh96}) that considering
also chemical instabilities resolves this
mystery.

\section{Phase Diagrams for Sonoluminescence}
To obtain phase diagrams of the bubble dynamics, one would have to solve
the full three dimensional gas dynamical PDEs inside the bubble,
coupled to the gas and fluid dynamics outside.
This is numerically intractable;
first, because the parameter space to be examined is huge, second, because
a solution has to be computed for many millions of forcing cycles to cover
the slow time scales of diffusional processes.

Consequently,
we do need approximations to continue. The most successful approximation
of bubble dynamics is the Rayleigh-Plesset (RP) equation 
for the radius
$R(t)$ of the bubble as a function of time.  This 
ODE was first written
down by  \cite{ray17}, while employed by 
the Royal Navy
to investigate
cavitation.  Modern improvements have been made by 
\cite{ple49,lau76,ple77,pro77},
and others. Recently,  \cite{fyr94} and
 \cite{loe95} were able to incorporate mass diffusion in
the RP approach.
This progress allowed us to study  diffusional
stability (\cite{bre96,hil96}).

The RP equation reads
\begin{eqnarray}
R \ddot R + {3\over 2} \dot R^2  &=&
{1\over \rho_{\rm w}} \left(p(R,t) - P(t) - P_0 \right)
   \nonumber \\
 &+& {R\over \rho_{\rm w} c_{\rm w}} {\D\over \D t}
     \left( p(R,t) -
    P(t)\right) - 4 \nu
    {\dot R \over R} -
    {2\sigma \over \rho_{\rm w} R}
\label{rp}
\end{eqnarray}
with the forcing pressure field 
\begineq
P(t)= P_{\rm a} \cos \omega t \enspace .
\label{poft}
\endeq
Parameter values for an argon bubble in water are
the surface tension
$\sigma = 0.073 {\rm kg}/{\rm s}^2$, water viscosity
$\nu = 10^{-6} {\rm m}^2/{\rm s}$, density
$\rho_{\rm w}= 1000 {\rm kg}/{\rm m}^3$, and
speed of sound $c_{\rm w}=1481 {\rm m}/{\rm s}$.
The driving frequency of the acoustic field is
$\omega/ 2\pi = 26.5 {\rm kHz}$
and the external pressure
$P_0= 1$\,atm.  We assume 
that
the pressure inside the bubble
varies according to the van der Waals law
\begineq
p(R(t)) = \left( {R_0^3 - h^3\over R^3(t) - h^3
}\right)^\kappa \enspace ,
\label{pressure}
\endeq
where
$h= R_0/8.86$
is the hard core van der Waals radius.
The effective polytropic exponent is taken to be $\kappa 
= 1$ (\cite{ple77})
for the small bubbles employed here.

\begin{figure}[htb]
\setlength{\unitlength}{1.0cm}
\begin{picture}(12,8)
\put(0.5,7.5){\LARGE a)}
\put(0.5,5){\LARGE b)}
\put(0.5,2.5){\LARGE c)}
\put(1.0,-0.5){\psfig{figure=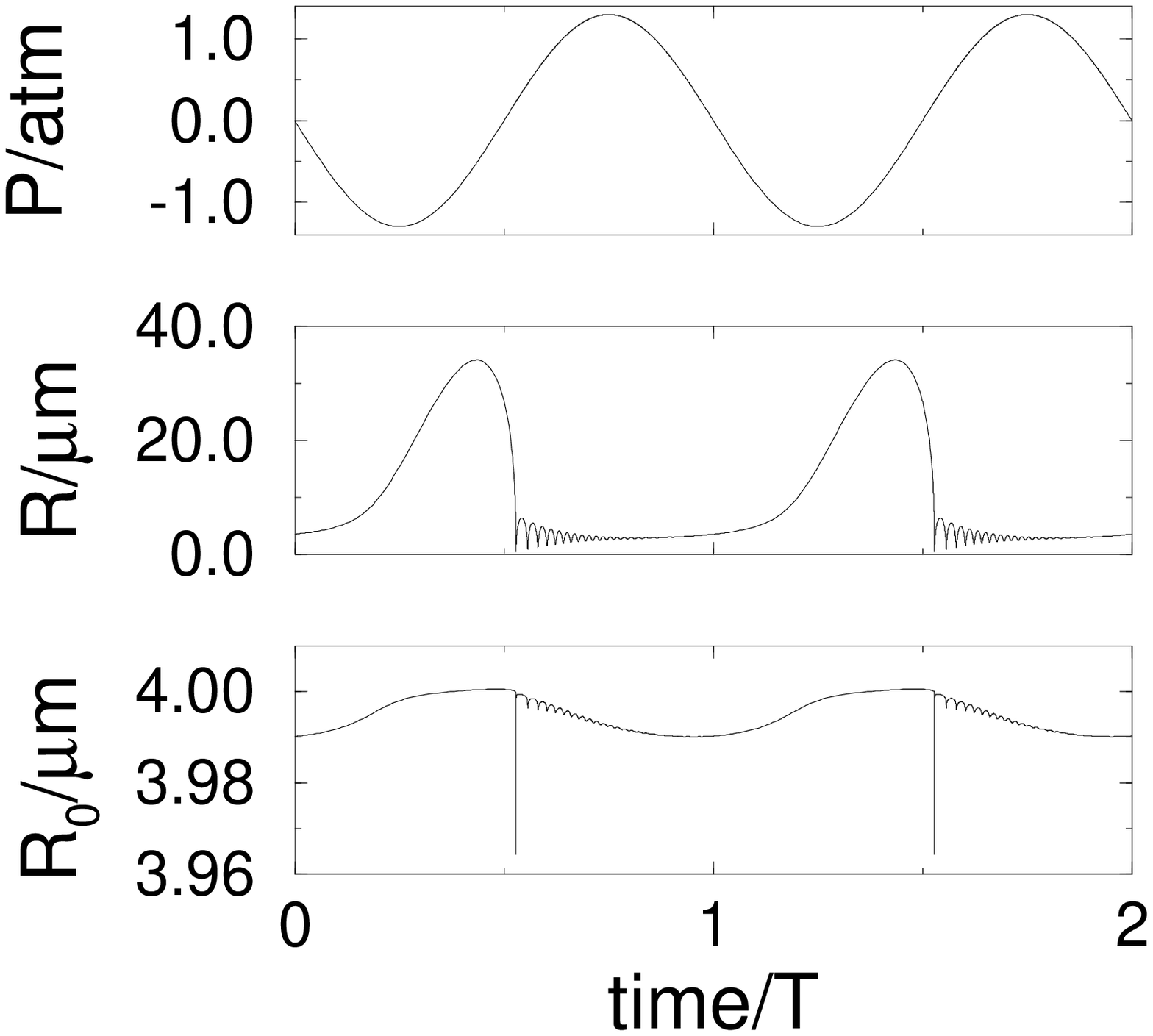,width=10cm}}
\end{picture}
\caption[]{
(a) Forcing pressure 
$P(t) = P_{\rm a} \cos \omega t $, $P_{\rm a}=1.3$\,atm for two 
cycles and
the corresponding (b) bubble radius $R(t)$ and (c) 
bubble ambient
radius $R_0(t)$.
The bubble is near an
equilibrium state. The gas concentration is 
$c^{\rm Ar}_\infty / c_0 = 0.004$. The ambient radius shown here
results from the numerical solution of the 
convection-diffusion PDEs,
for details we refer to \cite{hil96}.
}
\label{fig1}
\end{figure}

In figure \ref{fig1} we plot $R(t)$ resulting from
(\ref{rp}), together with the
forcing pressure $P(t)$. 
The radius shows a characteristic collapse which can
be very violent for large forcing $P_{\rm a}$. It is
right at this collapse when the light pulse is emitted. 
\cite{bar91} and \cite{loe93} 
have shown that the experimental curves of the radius look
very similar to that of figure \ref{fig1} and that (\ref{rp})
can be used to fit the dynamics of the radius. 
There may be small quantitative discrepancies with the
real dynamics, but they are clearly in the range of
the experimental error.

\begin{figure}[htb]
\centerline{\psfig{figure=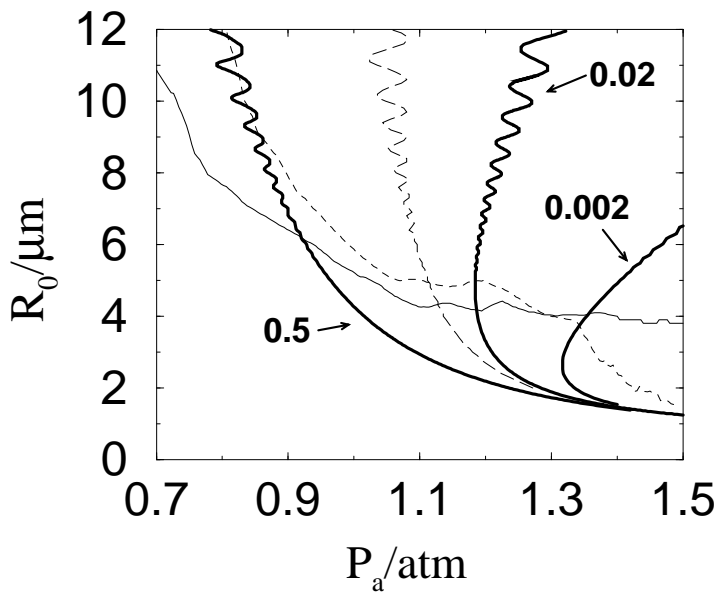,width=10cm}}
\caption[]{ 
Parametric instability (solid), Rayleigh-Taylor 
instability
(dashed), and shock condition (Mach=1, long 
dashed). SL is only possible in a small regime of large 
$P_{\rm a}\sim 1.3$\,atm
and small $R_0 \sim 3\mu{\rm m}$ where the bubble is 
spherically stable and
the shock condition is fulfilled. 
Also shown are the 
diffusive equilibria for $c^{\rm Ar}_\infty/c_0 =
0.5$, $ 0.02$, and $0.002$.
Whenever the slope is positive, the diffusive 
equilibrium is
stable. At $P_a=1.3$atm diffusively stable SL bubbles
are
only possible for very low concentration $c^{\rm Ar}_\infty/c_0 
\sim 0.002-0.004$. For larger $P_a$ even lower concentrations are
required.
}
\label{fig2}
\end{figure}

First, we address the question of 
{\it shape instabilities} ,
i.e., non-spherical deformations of the gas-water interface. 
There are basically
two different types of shape instabilities, 
distinguished by
the time scale over which they act.  {\it Parametric 
instabilities} act
over many bubble oscillation periods, and are the result 
of accumulation of
small perturbations.  {\it Rayleigh-Taylor
instabilities} occur when a light fluid is accelerated 
into 
a heavy fluid. Here, they happen during a short time directly after
the bubble collapse.  Although the process is very rapid
($\sim 10^{-9}$ \rm{s}),
detailed calculations (\cite{bre95,hil96}) show that at 
high
forcing, a perturbation of a few \AA\/ can grow to be as 
large as the bubble
during this time.  The differing time scales over which 
these
instabilities
act have consequences for the final state of the bubble:
after
a Rayleigh-Taylor instability, bubbles are typically 
destroyed, whereas after
a parametric instability, the bubble can remain trapped 
in the acoustic
field.
Figure \ref{fig2}
shows the borderline between stability and instability
in the ambient radius $R_0$ vs
 forcing amplitude $P_{\rm a}$ parameter space.
The diagram is calculated for an 
argon bubble in
water (\cite{hil96}).
Also depicted in figure \ref{fig2}
is the threshold curve beyond which the bubble wall speed exceeds
the speed of sound. We take this as a condition for the development
of shock waves
in the bubble. Following the common assumption that
shocks are necessary for sonoluminescence to occur
(\cite{gre93,wu93,mos94}), we thus get a
further restriction for the parameter range of sonoluminescing bubbles.

Let us now turn to {\it diffusive} stability.
Measurements of
the time between successive light flashes show that the
total mass of the bubble can remain constant to high 
accuracy
(\cite{gai90,bar91,bar95,loe95}) for days, if the
gas concentration is in an appropriate regime.  This result contradicts 
classical notions
about the dynamics of periodically forced bubbles:  An 
unforced bubble of
ambient radius $R_0$ {\it dissolves} over a diffusive 
time scale,
$\tau \sim \frac{\rho_0 R_0^2}{D(c_0-c_\infty)}$ 
(\cite{eps50}),
where $\rho_0$ is the
ambient gas mass density in the bubble and $D$ is the diffusion 
constant of the
gas in the liquid.
On the other hand, a strongly forced bubble grows
by rectified diffusion, as first discovered by \cite{bla49}
and later explored by 
\cite{ell69,cru80,cru94b}.
This is because
at large bubble radius the gas pressure in the bubble 
is small, resulting
in a mass flux into the bubble; conversely, when the 
bubble radius
is small there is a strong mass outflux. Both processes 
can be observed
in figure \ref{fig1}c which shows the ambient radius 
$R_0(t)$
corresponding to the actual bubble radius $R(t)$ of  
figure \ref{fig1}b.
Since the diffusive time scale is much larger
than the short time
the bubble spends at small radii,
gas cannot escape from the bubble during compression; 
the
net effect is bubble growth.
At a special value $R_0^{\rm e}$ of the ambient radius 
rectified diffusion and normal diffusion balance. 
The main result of \cite{bre96} was that contrary to
above intuition the equilibrium points $R_0^{\rm e}$
can be {\it stable}. \cite{hil96} showed that for {\it argon} bubbles
the  window of stability in
$c^{\rm Ar}_\infty/c_0$
is in quantitative agreement with experiment.

\begin{figure}[htb]
\setlength{\unitlength}{1.0cm}
\begin{picture}(12,8.5)
\put(0,-1){\psfig{figure=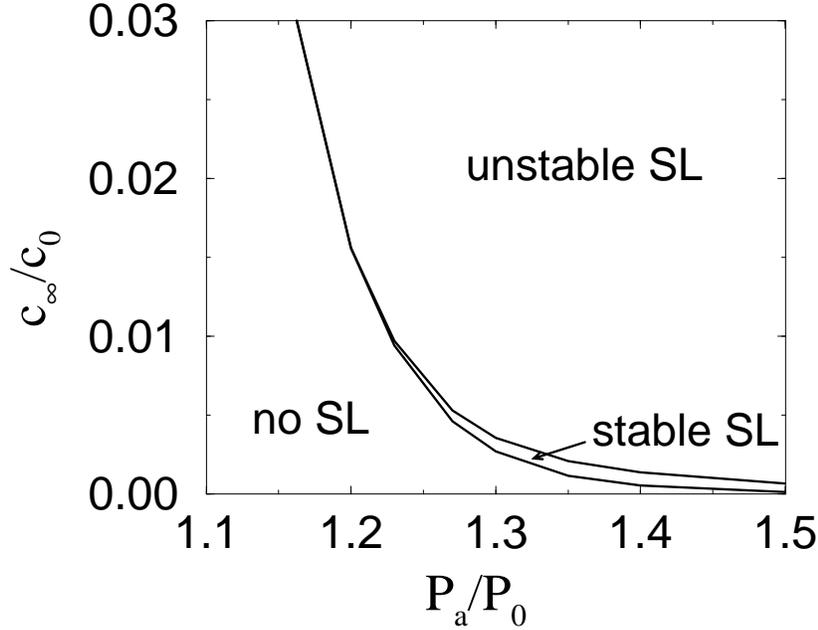,width=11cm}}
\end{picture}
\caption[]{ 
Phase diagram in the $c^{\rm Ar}_\infty/c_0$ vs $P_{\rm a}/P_0$ 
parameter space.
For the SL phase we assumed that the light production
mechanism works beyond the Mach=1 curve.
Bubbles at very large $P_{\rm a}$ may be Rayleigh-Taylor unstable.
}
\label{fig3}
\end{figure}

\begin{figure}[p]
\setlength{\unitlength}{1.0cm}
\begin{picture}(12,14.5)
\put(0.4,13.7){\LARGE a)}
\put(0.4,6){\LARGE b)}
\put(0.9,6.8){\psfig{figure=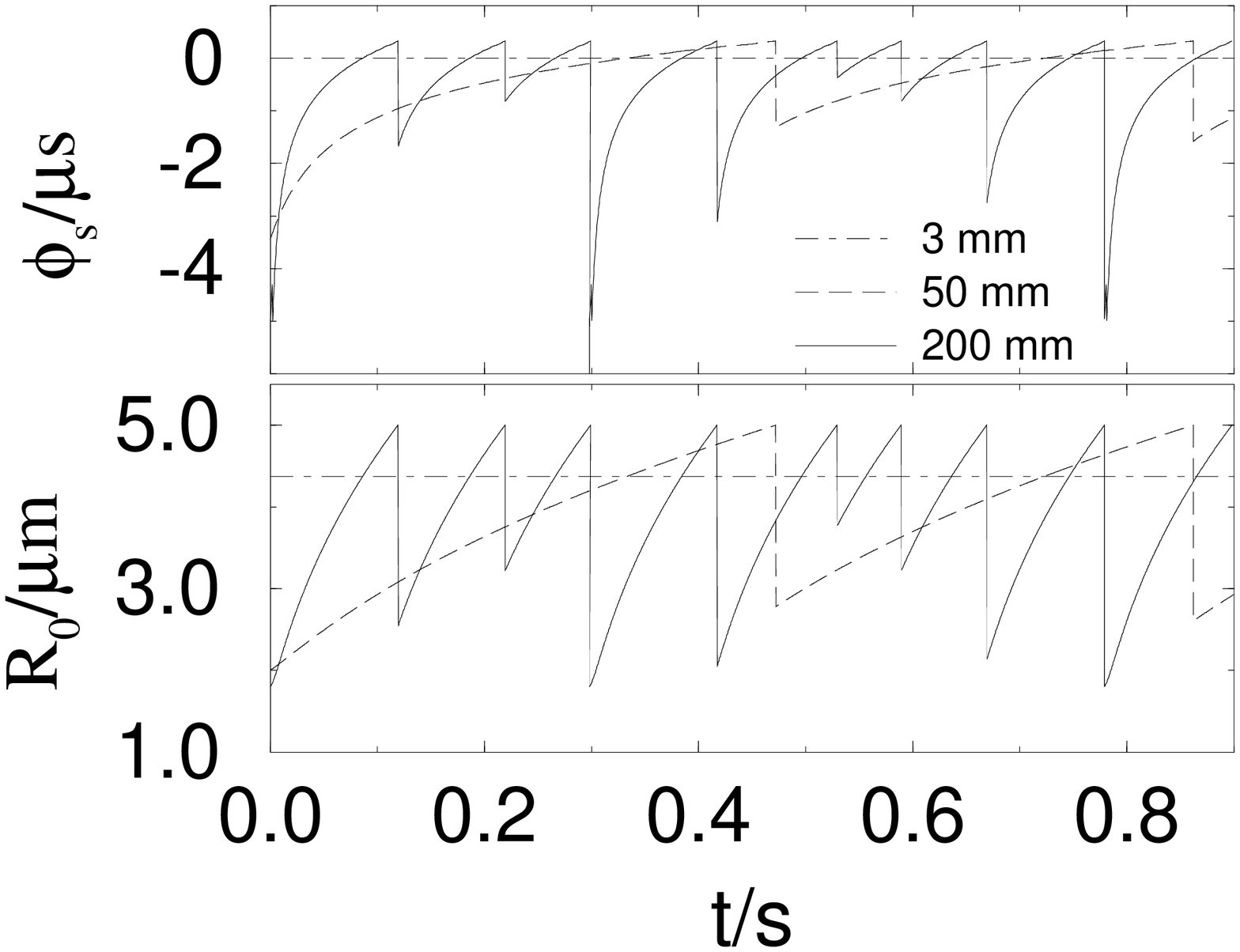,width=9cm}}
\put(1.55,-0.4){\psfig{figure=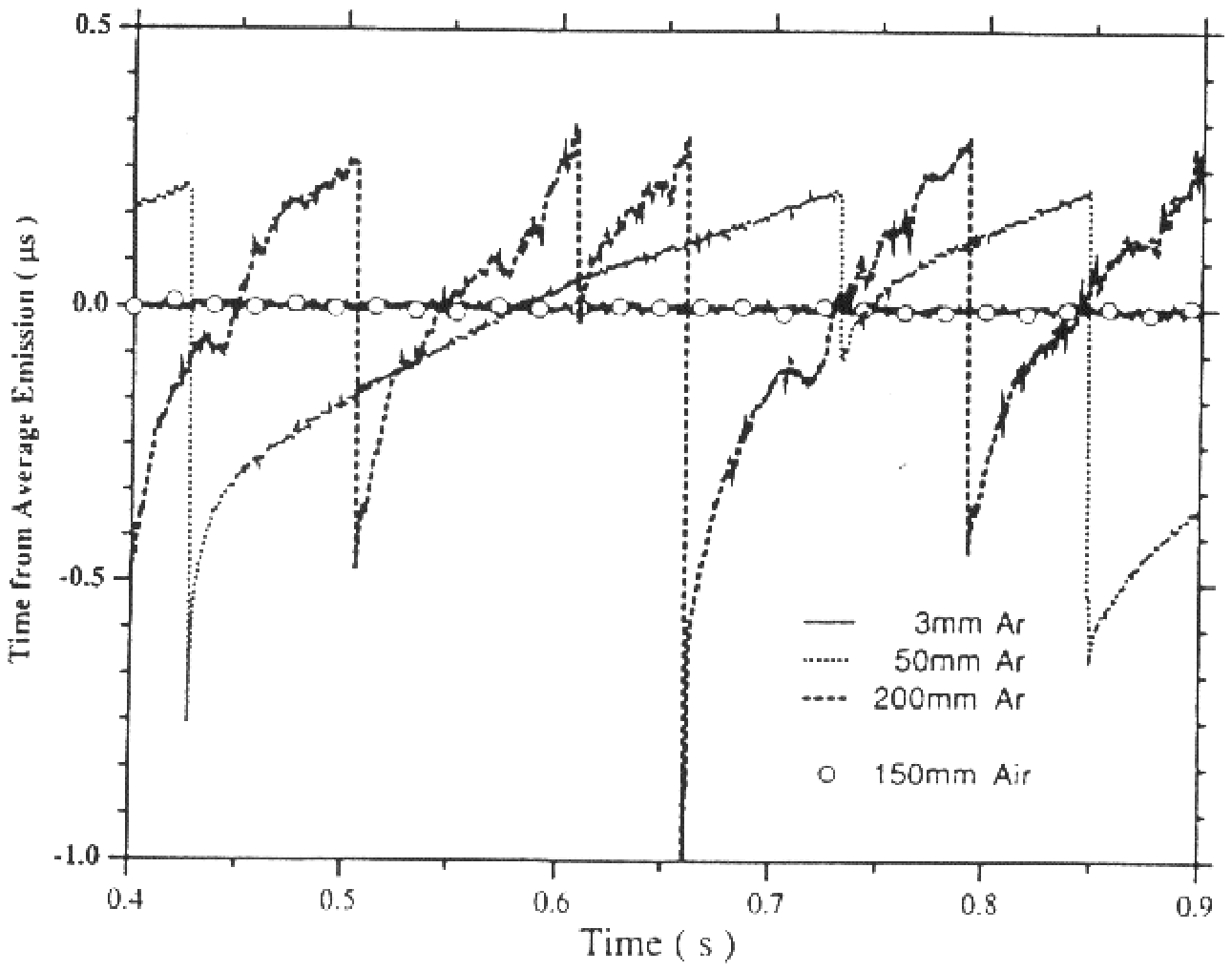,width=9cm,angle=-0.6}}
\end{picture}
\caption[]{
(a) The phase of light emission $\phi_{\rm s}(t)$ 
(upper)
and the corresponding
ambient radius $R_0(t)$  (lower) for $P_{\rm a}= 
1.3$\,atm
for three different gas 
concentrations
$c^{\rm Ar}_\infty / c_0 =  0.004$,
$c^{\rm Ar}_\infty / c_0 =  0.066$, and 
$c^{\rm Ar}_\infty / c_0 =  0.26$, corresponding to a gas 
pressure of
$3$\,mmHg, $50$\,mmHg, and $200$\,mmHg, respectively.
The strength of the micro-bubble pinch-off at $R_0=5\mu 
m$, i.e.
the decrease of the ambient radius, is picked randomly.
The concentration  values are
chosen as in 
the experiment of \cite{bar95} which is shown in
(b), copied from Fig.\,4 of that reference.
That figure also shows the relative phase of light emission
for {\it air} bubbles: Stable SL is achieved at {\it much higher
 concentration}
$c^{\rm air}_\infty / c_0 =  0.2$, corresponding to a gas 
pressure of 150\,mmHg. It is this discrepancy
between argon (or other inert gases)
and air which we resolve in this paper.
}
\label{fig4}
\end{figure}

\begin{figure}[htb]
\centerline{\psfig{figure=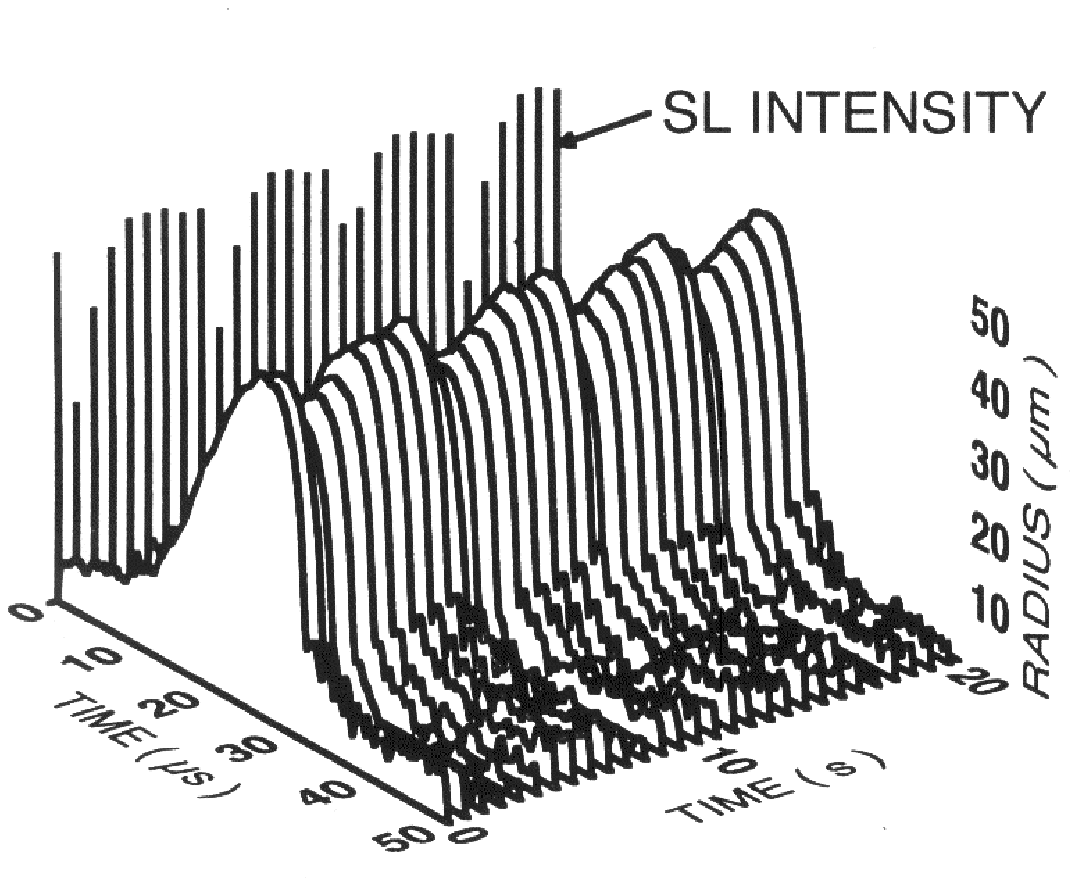,width=8cm}}
\caption[]{
This figure is reproduced from figure 6 of \cite{loe95}.
It shows the
SL intensity and bubble radius as a function of time for 
a nitrogen
bubble with $5\%$ argon.
The drive level $P_{\rm a}$ is near the upper threshold of SL.
The gas is dissolved at $150$\,mmHg, i.e.
$c^{\rm mix}_\infty/c_0\approx 20\%$.
The bubble radius exhibits the characteristic
pattern of growth and microbubble pinch-off of unstable SL.
}\label{fig5}
\end{figure}

Figure \ref{fig3} shows the total phase diagram as a 
function of
$c^{\rm Ar}_\infty/c_0$ and $P_{\rm a}/P_0$. Three different states are 
possible:
The label {\it stable SL} denotes the region of 
parameter space where
the bubble can be diffusively stable and also stable to shape 
oscillations.
In the region
{\it no SL} the bubble dissolves, while in
the {\it unstable SL} domain the bubble is not in
diffusive equilibrium but can nonetheless survive.
That this is possible is quite
nontrivial, and was first
shown by experiments of \cite{gai90}
 and \cite{bar94}.

In the unstable SL phase the bubble is growing by
rectified diffusion until it hits the parametric instability.
The bubble becomes shape unstable and microbubbles pinch off.
If the remaining bubble is still large enough, the process
begins again. 
Figure \ref{fig4}a
 shows the ambient radius
 $R_0$ and the
 corresponding phase $\phi_{\rm s}$ of the light emission relative 
to the forcing
field as a 
function of time as calculated from 
our theoretical Rayleigh-Plesset
approach (\cite{hil96}).
We chose the same three gas concentrations as in
the experiment of \cite{bar95} which we
present in Fig.\,\ref{fig4}b for comparison.
Both in our calculations and in experiment
we are in the
unstable SL regime
for the two larger argon partial pressures (200\,mmHg and 50\,mmHg), whereas
the lowest value of 3\,mmHg is in the stable SL regime, again
both in theory and experiment. Argon bubbles thus seem to be well understood.

Another experimental result
(now for a gas mixture)
in the unstable SL regime is shown in
Fig.\,\ref{fig5}.
We observe the maximal radius $R_{\rm max}$ and the light intensity
growing on the diffusional time scale $\sim 1{\rm s}$ until the bubble hits
the parametric instability and the process starts over. Note that
our interpretation of the unstable SL regime also
accounts for the observed ``dancing'' of the bubble: the pinch-off of the
microbubbles leads to recoils of the bubble. 
The dancing frequency should be the higher the larger $c^{\rm Ar}_\infty $ is.

Although the above results for pure argon bubbles
are in excellent agreement 
with experiments,
we still have above mentioned severe discrepancy for 
air bubbles and other gas mixtures that stable SL is observed
for much larger concentrations than expected from our phase diagram
Fig.\,\ref{fig3}. E.g., in air 
we have stable SL for 
$c^{\rm air}_\infty/c_0 \approx 0.1-0.2$, depending on the forcing
pressure.  This discrepancy
was first pointed out in the qualitative arguments of 
\cite{loe95} who found that air bubbles in 
the range where
stability is observed should be unstable from a theoretical
point of view, based on RP dynamics.  The 
reason for this can be seen
in the phase diagram above: 
if the diagram is repeated for air instead of argon, the
stable equilibria 
still occur at very low concentrations, namely 
$c^{\rm air}_\infty/c_0 \approx 0.004.$
Based on this discrepancy, L\"ofstedt et al. 
hypothesize 
an ``as yet unidentified
mass ejection mechanism.''

\section{A Chemical Resolution of the Paradox}
In joint work 
with Todd F. Dupont and Blaine Johnston of the University 
of Chicago (\cite{loh96}), we have
pointed out that this paradox can be resolved by considering
{\it chemical instabilities}.
Namely, it is well
known 
that the temperature in the bubble becomes very high 
after the collapse. Fits
of the spectra of emitted radiation to a 
black body law give
effective temperatures of up to $25000 {\rm K}$ (\cite{hil92}).  
This
temperature exceeds the dissociation temperature for 
nitrogen gas
($\sim 9000{\rm K}$) as well as for all other molecular 
constituents of air.
The nitrogen, oxygen, and hydrogen radicals will recombine to finally
form NO, NO$_2$, and/or NH$_3$.
All of these gases are highly soluble in water, forming nitric
and nitrous acid and NH$_4$OH. The consequence is that the bubble is finally
depleted from the molecular constituents of air.

Even beyond ``burning'' of all initial nitrogen in the bubble we expect
that whenever the bubble is large it will still suck gas from the water into the
bubble which will also be ``burnt'' during the next compression cycle.
The bubble thus constitutes a reaction chamber for the dissolved
molecular gas. The reaction products should be detectable. \cite{loh96}
estimated a production rate of $\sim 10^{11} {\rm mol}/{\rm s}$ per container
volume ($\sim 1{\rm l}$). Let us assume that the main product is NH$_4$OH.
Starting with the optimal pH=7 we should thus get a doubling of OH$^-$ ions
within about three hours and a pH change up to pH=8 (i.e., $c_{\rm{OH}^-} =
10^{-6}{\rm mol}/{\rm l}$) in about a day of running the experiment.
Many other ways of detection rather
than simply measuring the pH seem possible. An ion chromatographer will give a
detailed answer on what ions are produced.

For the present paper we only need the assumption that even when air was
initially dissolved in water, a sonoluminescing bubble essentially consists
of inert gases as
these are the only gases which do not react with water at the high temperatures
achieved in single bubble SL; they are simply recollected by the expanding
bubble as seen in Fig.\,\ref{fig1}c.

The dissociation hypothesis suggests that for gas mixtures
the relevant concentration quantity is still the {\it argon} concentration
$ c_\infty^{\rm Ar}/c_0$ 
which now is 
\begineq
{c_\infty^{\rm Ar}\over c_0 }
 =q \cdot  \frac{c_\infty^{\rm mix}}{c_0} \enspace ,
\label{psi}
\endeq
where $q$ is the percentage of argon in the mixture.
For air we have $q=0.01=1\%$. Here we have neglected the differences
in solubility $c_0$ of the different gases.
This difference enters in two counteracting ways: first,
(\ref{psi}) must be corrected by the ratio
$c_0^{\rm Ar}/c_0^{{\rm N}_2}$
of the solubilities between argon and nitrogen,
which is about three. But second,
because of the different solubilities the percentage of argon in
{\it dissolved} air will not be the usual $1\%$, but will be 
larger. Possibly, this percentage even depends on the degree of degassing
of the water.
These two effects will probably roughly compensate each other
and we  decided to disregard
them for the time being.

In the following section we discuss how the dissociation  hypothesis 
can explain the experimental UCLA 
results on gas mixtures (\cite{hil94,bar94,bar95,loe95}).
We point out that our detailed unpublished 
investigations of
mixtures 
of gases demonstrate that it is impossible to understand 
these discrepancies
within diffusive theories alone, i.e., by disregarding 
chemical reactions.

\begin{figure}[p]
\setlength{\unitlength}{1.0cm}
\begin{picture}(12,15.2)
\put(1.5,6.8){\psfig{figure=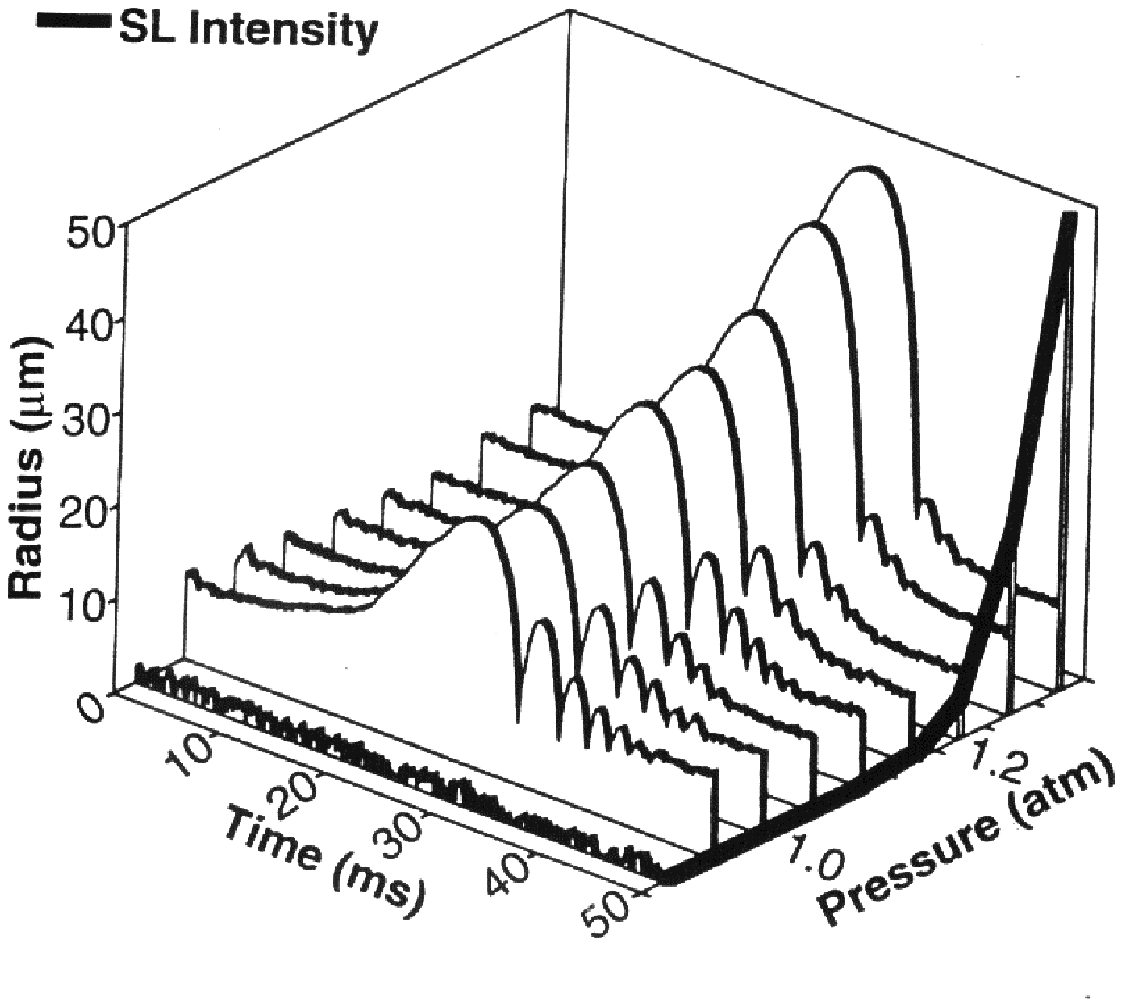,width=8.8cm,angle=1.5}}
\put(1.4,-0.3){\psfig{figure=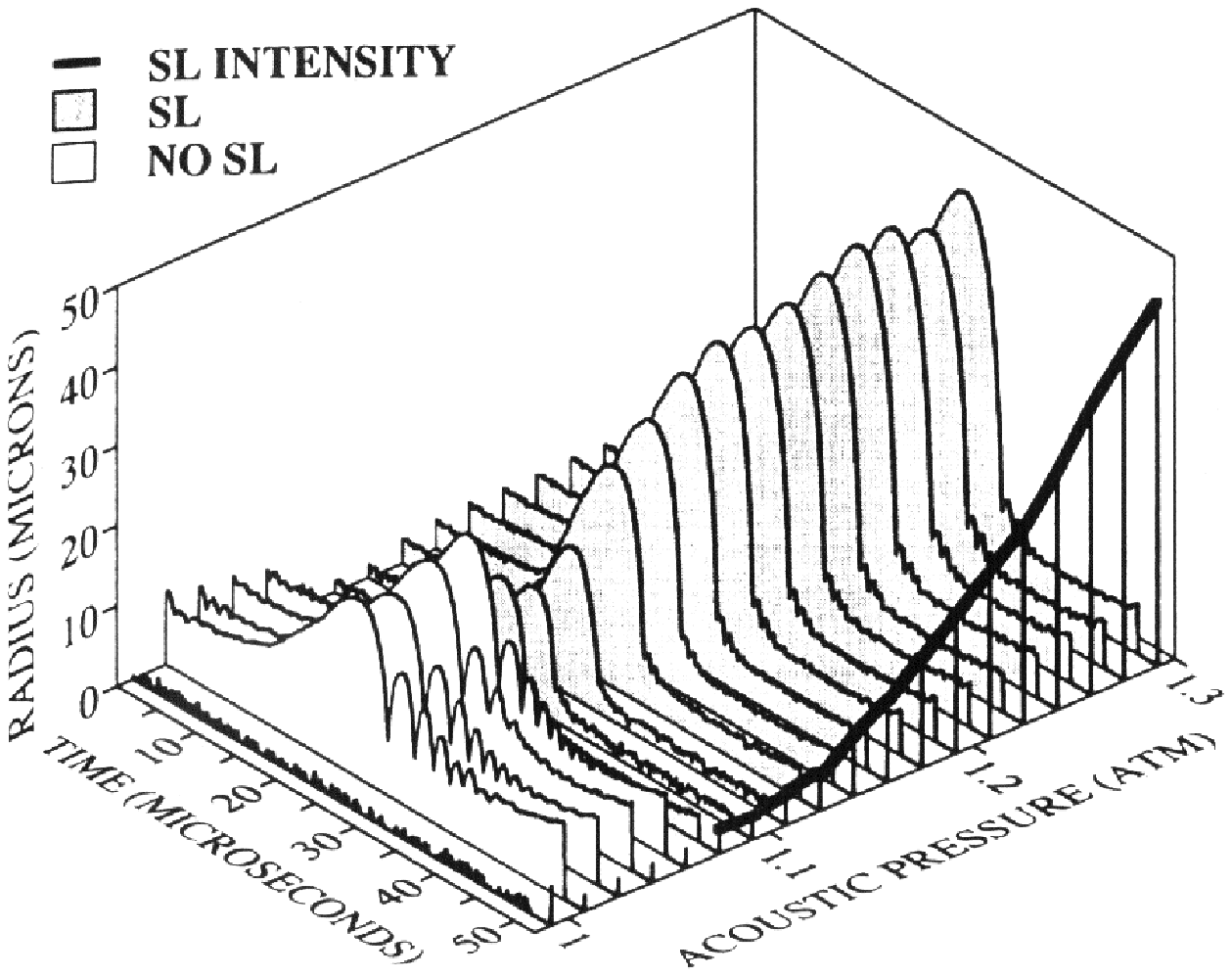,width=9cm}}
\put(0.4,14.2){\LARGE a)}
\put(0.4,6){\LARGE b)}
\end{picture}
\caption[]{
Transition towards the SL regime for argon bubbles (a) 
and for air bubbles
(b). Only for air bubbles a breakdown in the radius is 
seen
near the onset of SL, signaling the threshold for 
nitrogen dissociation.
Figure (a) is a reproduction from figure 4 from 
\cite{hil94}, the total gas concentration is about 
150\,mmHg.
This means $c^{\rm air}_\infty/c_0 = 0.2$ in the unstable SL 
regime in
agreement with the observations.
Figure (b) is a reproduction from figure 2 of \cite{bar94}.
The gas saturation is about $10\%$, corresponding to
$c^{\rm Ar}_\infty/c_0 = 0.01 \cdot 0.1 = 0.001$.
According to Fig.\,\ref{fig3} we have stable SL around $P_{\rm a}
\sim 1.4$\,atm
which again is in rough agreement with experiment.
}\label{fig7}
\end{figure}

\section{Comparison with the UCLA Experiments}
Where to expect stable SL in air bubbles?
From our phase diagram Fig.\,\ref{fig3} we know that
for $P_{\rm a}=1.3$\,atm stable SL is found in a small window of 
$c^{\rm Ar}_\infty/c_0 \sim 0.3 - 0.4\%$ for pure argon gas.
Equation (\ref{psi}) thus gives
$c^{\rm air}_\infty/c_0 \sim 30 - 40\%$ for air dissolved in water.
The experimental values seem to be slightly smaller, but
clearly within the error range of both our assumptions and
the measurements.

The transition towards SL with increasing forcing pressure $P_{\rm a}$ is
shown in Fig.\,\ref{fig7} for both argon and air bubbles.
For pure argon bubbles the transition to SL is very smooth.
For air, however,
one can observe a breakdown of the bubble radius at about
$1.1$\,atm, signaling that the dissociation threshold of
N$_2$ is achieved. 
Before the transition
the bubble is filled with a 
mixture of nitrogen, oxygen, and argon,
and the 
ambient radius is determined by the combination of all 
three gases.  After
the dissociation threshold, 
only argon is left in the bubble.

We now also see why stable SL is so easily obtained in
air bubbles: The window of stability is larger compared to
argon bubbles and the water has to be degassed much less.
But even more friendly conditions are possible:
Our theory suggests that it should be possible to obtain stable
SL {\it without} degassing, namely by choosing the percentage
$q$ of argon so that the window of stable SL is around
$c^{\rm mix}_\infty/c_0
\sim 100\%$. With $q\sim 0.35\%$ we obtain as window of stable SL
$c^{\rm mix}_\infty/c_0\sim 85 - 115\%$.

\begin{figure}[htb]
\centerline{\psfig{figure=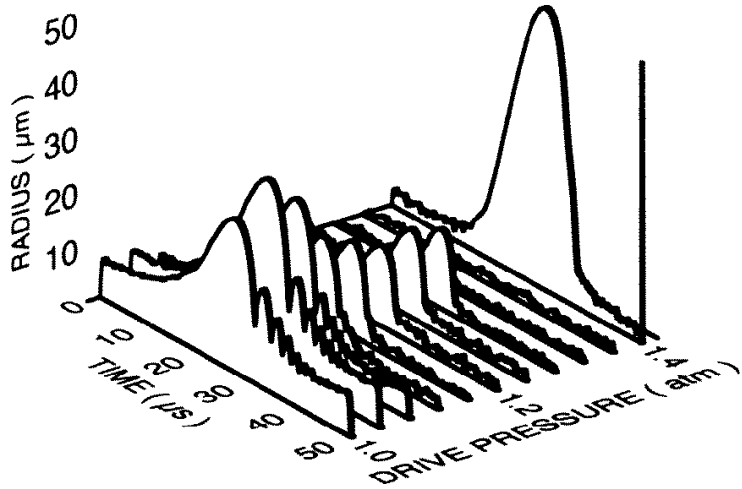,width=10cm}}
\caption[]{
This figure is reproduced from figure 11 of \cite{loe95}.
It shows the transition to SL for a bubble filled with an initial $0.1\%$
xenon in nitrogen gas mixture at a
partial pressure of 150\,mmHg. According to our 
dissociation hypothesis this
corresponds to $c^{\rm Xe}_\infty/c_0 = 0.0002$ for a pure sonoluminescing
xenon bubble.
From our phase diagram Fig.\,\ref{fig3} which is (with tiny
corrections) also valid for xenon we conclude
that bubbles driven at $P_{\rm a}=1.3$\,atm dissolve for these 
low concentrations,
whereas bubbles at $P_{\rm a}=1.4$\,atm show stable SL, just as 
seen in
this experimental figure.
}\label{fig6}
\end{figure}

For even lower concentration $q<0.3\%$ (at $P_{\rm a}=1.3$\,atm)
no stable SL regime
is left for $c_\infty^{\rm mix}/c_0 < 100\%$.
The experiment of \cite{loe95} with $q=0.1\%$ xenon in nitrogen at
$c_\infty^{\rm mix}/c_0=0.2$ shown in
Fig.\,\ref{fig6} relates to
exactly that
situation.  \cite{loe95} find a range 
of forcing
pressures $P_{\rm a} \approx 1.3-1.4$\,atm where
stable bubbles cannot be seeded.  Above
$1.4$\,atm, stable SL exists, and below $1.3$\,atm there are 
non--sonoluminescing
bubbles.  The 
reason for this ``gap''
is that at these high forcing pressures 
all nitrogen is dissociated and the bubble only contains the inert
gas xenon.
Then, according to our phase
diagram Fig.\,\ref{fig3} (which also holds for xenon), 
at $P_{\rm a}=1.3$\,atm the xenon concentration $c_\infty^{\rm Xe} =
q\cdot c^{\rm mix}_\infty/c_0 = 0.001\cdot 0.2 = 0.0002$ is in the {\it no SL}
regime and bubbles dissolve. 
But at higher 
forcing pressure $P_{\rm a}\sim 1.4$\,atm we can again enter the {\it stable SL}
regime for that concentration, see Fig.\,\ref{fig3}: a pure xenon
bubble can emit light and acts as a chemical reaction chamber, transferring
the dissolved nitrogen to its high temperature reaction products.

\begin{figure}[htb]
\centerline{\psfig{figure=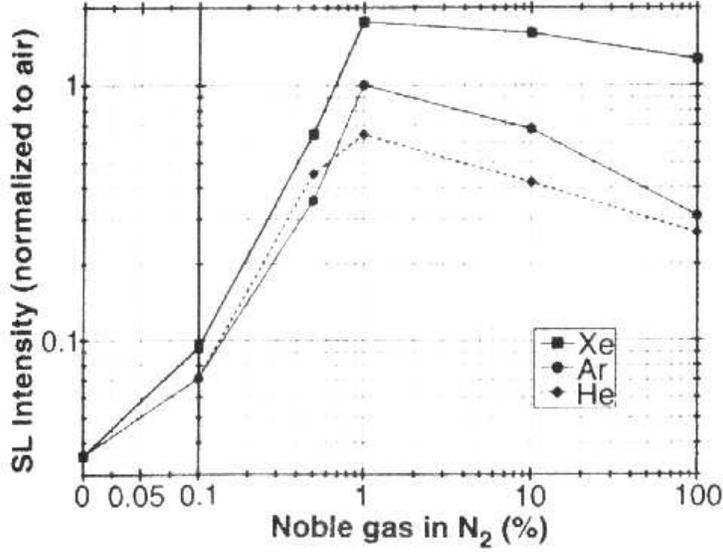,width=10cm}}
\caption[]{
This figure is reproduced from figure 1 of \cite{hil94}.
It shows the SL intensity from a SL bubble in water as a 
function
of the percentage (mole fraction) of noble gas mixed 
with nitrogen.
The gas mixture was dissolved in water at a pressure 
head of 150\,mmHg, i.e., $c_\infty^{\rm mix} /c_0 \sim 0.2$.
}\label{fig8}
\end{figure}

The transition from the no SL regime in Fig.\,\ref{fig3}
to the SL regime
can also be seen
in Fig.\,\ref{fig8}, where the SL light intensity is plotted
as a function of $q$ for 
fixed $P_{\rm a}$ (we assume
$P_{\rm a}=1.3$\,atm) and fixed $c^{\rm mix}_\infty/c_0\sim 0.20$.
According to our theory the transition for
$c_\infty^{\rm Ar}/ c_0 =0.003$
should occur at $q=
(c_\infty^{\rm Ar}/ c_0) 
/(c^{\rm mix}_\infty/c_0) = 1.5\%$ in pretty good
agreement with Fig.\,\ref{fig8} where we indeed observe that
strong SL is ``switched on'' at about that concentration.
Our theory also predicts that near the switch on we always
have stable SL, whereas for larger $q$ unstable SL develops.
Here we expect unstable
SL for $q>2\%$, a prediction which should be verified.

An example for unstable SL was already shown in Fig.\,\ref{fig5}.
Indeed, for that figure we have $q=5\%$. Thus, from 
(\ref{psi}) we have $c_\infty^{\rm Ar}/c_0 = 0.05\cdot 0.2=0.01$ and 
according to our phase diagram Fig.\,\ref{fig3} and
equation (\ref{psi}) we are well in the unstable SL regime, just in
agreement with the observations.

Next, we consider the case of a pure nitrogen bubble. 
 Below the dissociation
threshold it 
is ``jiggling'', i.e. growing by rectified diffusion and 
shedding
microbubbles. 
Above the threshold, the nitrogen is taken away, so the 
bubble will
necessarily dissolve.  
Indeed, \cite{hil94} show that pure nitrogen 
bubbles
dissolve in a 
matter of seconds.  In figure 5 of their work
it is demonstrated that pure nitrogen bubbles show
``periodic''
oscillations in their light intensity.  We do not 
understand what sets the
oscillation period, 
though we speculate that it is connected to the 
dissociation of nitrogen
in the bubble, as 
well as the unstable ``jiggling'' of the bubble.  
Perhaps by
understanding the 
origin of these oscillations one could deduce the 
temperature inside
the bubble.

Another experiment supports our  dissociation 
hypothesis.
\cite{hil95} analyze SL in 
$\rm{H}_2$ and
$\rm{D}_2$ gas bubbles, both in normal and in heavy water. 
One would expect that the SL intensity curves would 
group
according to the gas content as it is the gas dynamics 
inside the
bubble which determines the strength of the light 
emission. 
However the four
(H$_2$ in H$_2$O,
 H$_2$ in D$_2$O,
 D$_2$ in H$_2$O, and
 D$_2$ in D$_2$O)
experiments group according to the 
surrounding
liquids. This suggests the following scenario:
Both the gas and the liquid vapor in the bubble
dissociate during the (hot) compression phase and 
recombine later on during
expansion. As much more liquid than gas is around,
recombination will yield gases primarily composed of
the constituents of the liquid vapor.
E.g., in the experiment using $\rm{H}_2$ in $\rm{D}_2\rm{O}$, 
the $\rm{H}_2$ 
will shortly be replaced by $\rm{D}_2$ formed from the heavy water,
whereas the amount of $\rm{H}_2\rm{O}$ formed from $\rm{H}_2$
will be negligible. Thus the density of the bubbles in
D$_2$O  will eventually be larger  by a factor of two compared
to the density of bubbles in H$_2$O.
Following our 
recently proposed laser theory of SL (with R.\ Rosales, (\cite{bre96d})),
it is this density difference
which may be the origin of the 
considerably different SL intensities 
observed in normal and heavy water by \cite{hil95}.

\begin{figure}[htb]
\centerline{\psfig{figure=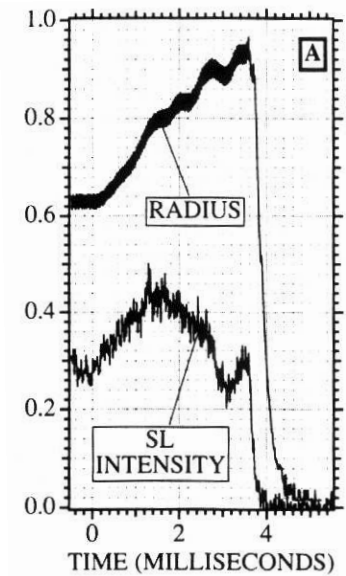,width=5cm,angle=1}}
\caption[]{
This figure is reproduced from figure 4 of \cite{bar94}.
It shows the dynamic response of the radius and the 
intensity
of SL to a sudden change in the drive level $P_{\rm a}$. The 
bubble is
boosted into the unstable SL regime where it finally 
becomes
shape unstable and bursts.
}\label{fig9}
\end{figure}

There is one final experimental figure which we want
to discuss here. It is Fig.\,\ref{fig9}, showing the
{\it maximal} radius
$R_{\rm max}(t) = {\rm max} \{R(t')| t\le t'\le t+T \}$
and
the intensity as a function of time when the forcing pressure
$P_{\rm a}$ is suddenly strongly increased. 
On this increase, the bubble is pushed in the unstable SL domain and
starts to grow rapidly thanks to rectified
diffusion. But why is the growth rate of $R_{\rm max}$ so wiggly
whereas the growth rate of the ambient radius does not show
these wiggles, as seen from Fig.\,\ref{fig4}?

\begin{figure}[htb]
\setlength{\unitlength}{1.0cm}
\begin{picture}(12,8)
\put(0,-0.7){\psfig{figure=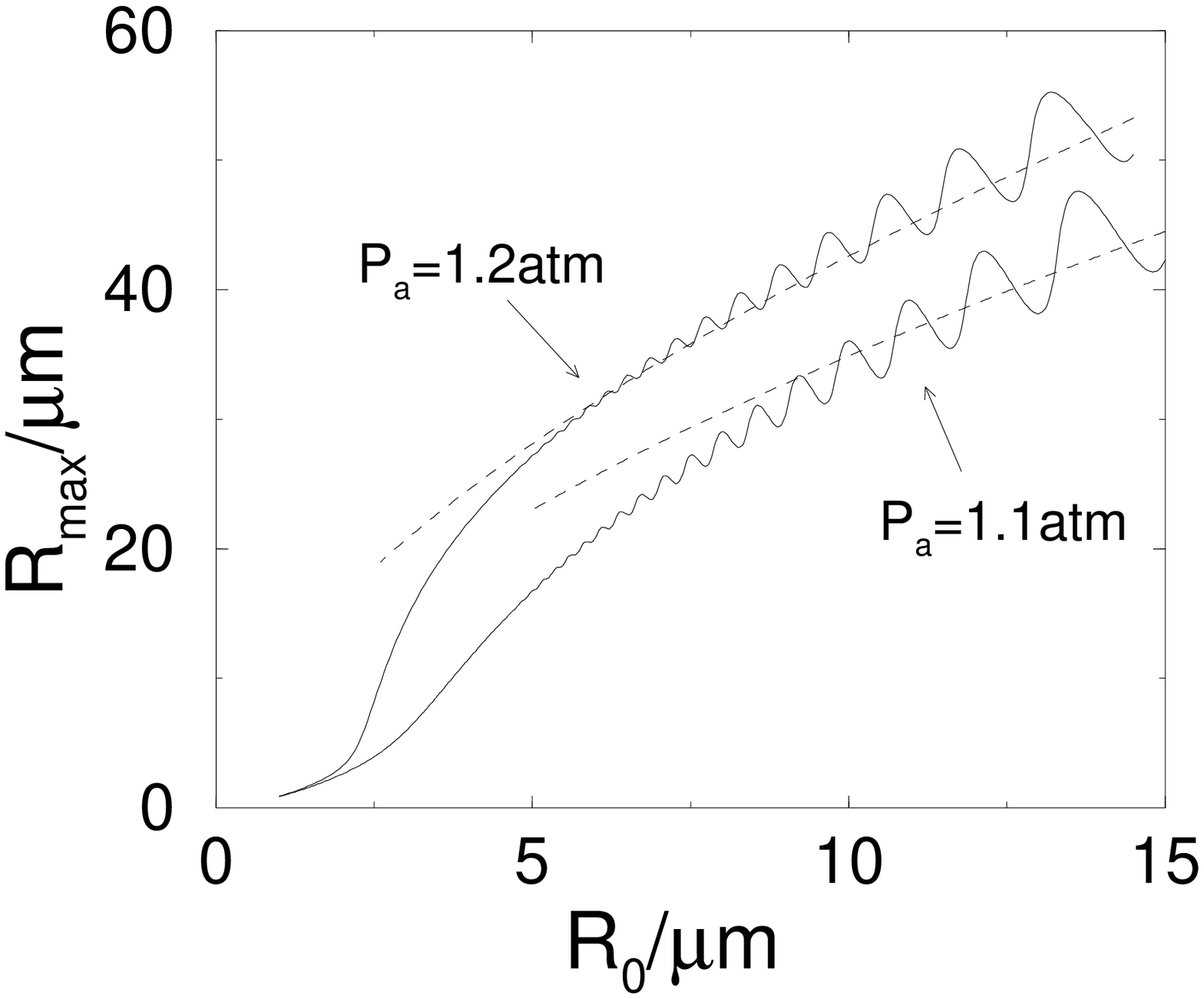,width=10cm}}
\end{picture}
\caption[]{
The maximal bubble radius
$R_{\rm max}$ as a function of the ambient radius $R_0$
for two forcing pressures $P_{\rm a}=1.1$\,atm and 1.2atm.
It shows characteristic wiggles. The dashed lines depict 
the
$R_{\rm max} \propto R_0^{3/5}$ scaling law.
}\label{fig10}
\end{figure}

The answer is that $R_{\rm max}$ itself shows wiggles as a function
of the ambient radius, as observed in \cite{bre96}
and analyzed in detail by \cite{bre96c}. In figure \ref{fig10} we
present $R_{\rm max}(R_0)$, whose wiggles
are probed in figure \ref{fig9} by the growing $R_0$.
The averaged scaling law is given by
$R_{\rm max} \propto R_0^{3/5}$ (\cite{bre96c}).

In diffusionally stable regimes these wiggles may lead to
multiple stable equilibria, as suggested in \cite{bre96}
and possibly observed by \cite{cru94b}.
\cite{bre96c} pointed out that the origin of the wiggles and the resulting 
multiple equilibria are resonances in the RP dynamics.

\section{Conclusions}
This paper 
has summarized our current understanding of the bubble
dynamics leading to single bubble
sonoluminescence.
Our central theoretical result is the phase diagram of
SL figure \ref{fig3} from \cite{hil96}.  
It can only be calculated in the Rayleigh-Plesset
SL bubble approach including both shape oscillations and diffusion.
The phase diagram allows to quantify the differences
between argon and air bubbles found by \cite{loe95},
guiding them to suggest a ``yet unidentified
mass ejection mechanism'' for air bubbles. We now believe
that this mechanism is chemical, namely the dissociation
of gas molecules in highly forced gas bubbles and their
subsequent reactions. As pointed out in this paper, with this 
``nitrogen--dissociation hypothesis'' many 
experimental results can be
explained.
To further verify this hypothesis,
experiments should search for the various reaction 
products.

One issue 
that we have barely touched in this paper is the 
mechanism for the light
emission.  The 
traditional theory is that shocks are emitted from the 
bubble wall, and
heat up the gas in the 
bubble to such an extent that light emission is induced.
Experimentally
there are strong dependences of the light intensity on
liquid, gas and forcing
pressure.  These cannot be explained within the context of 
simple bubble dynamics presented 
here, and also do not seem to be understandable within the shock
theory.  The shock theory relies on 
a singular solution to the hydrodynamic equations;  singularities
are typically of universal character and very insensitive to external
conditions.  However, the fact
that experiments on single bubble SL demonstrate such
sensitivity to parameters 
contradicts this principle.
The most dramatic example of this is the 
initial observation of single bubble
SL by \cite{gai90}:  it caused great excitement 
because the light intensity is so much higher
than in multibubble SL.  However, the shock theory should
apply equally well to both cases.
These ideas suggest that the shock theory has  intrinsic shortcomings.
To overcome these problems \cite{bre96d} recently
proposed a ``hydrodynamic laser'' theory for SL bubbles.
It stipulates that
the bubble can accumulate acoustic energy inside the bubble which
acts as a kind of acoustic resonator.

We hope that our present and ongoing work  will help to
answer one of the essential remaining
questions on SL: How hot does the gas become
in the center of the bubble?
Our hope is that the presence of intermediate or final
chemical reaction products can be thought of as a thermometer.
-- The perspectives are appealing: A tiny air
bubble in water is used as a
chemical reaction chamber
and a micro-laboratory for high temperature chemistry!

\vspace{1cm}

\noindent {\bf Acknowledgements:}\\
We thank our colleagues and collaborators B.\ Barber, L.\ Crum, 
K.\ Drese, T.\ F.\ Dupont, B.\ Gompf, S.\ Grossmann, 
B.\ Johnston,
L.\ Kadanoff, W.\ Kang, D.\ Oxtoby, Th.\ Peter,
S.\ Putterman, R.\ Rosales and J.\ Young
for
many useful and 
thought provoking discussions and suggestions.  This research was 
supported by the DFG
through its SFB185.
MB acknowledges an NSF postdoctoral fellowship.


\begin{thebibliography}

\bibitem{}{bar91}{Barber and Putterman (1991)}
{\sc Barber, B.~P.,} and  {\sc Putterman, S.~J.,}
``Observation of synchronous picosecond sonoluminescence'',
 Nature (London) {\bf 352},  318-320  (1991);
``Ligth scattering measurements of the repetitive supersonic implosion
of a sonoluminescing bubble'',
Phys. Rev. Lett. {\bf 69},  3839-3842  (1992).

\bibitem{}{bar94}{Barber et al. (1994)}
{\sc Barber, B.~P.,  Wu, C. C.,   L\"ofstedt, R.,  Roberts, P. H.,}
 and  {\sc Putterman, S. J.,}
``Sensitivity of sonoluminescence to experimental parameters''
Phys. Rev. Lett. {\bf 72},  1380-1383  (1994).

\bibitem{}{bar95}{Barber et al. (1995)}
{\sc  Barber, B.~P., Weninger, K.,  L\"ofstedt, R.,}
 and  {\sc Putterman, S.~J.,}
``Observation of a new phase of sonoluminescence at low partial pressures'',
 Phys. Rev. Lett.
  {\bf 74},  5276-5279  (1995).

\bibitem{}{bla49}{Blake (1949)}
{\sc  Blake, F.~G.,} Harvard University Acoustic Research Laboratory Technical
  Memorandum {\bf 12},  1  (1949).


\bibitem{}{bre95}{Brenner, Lohse, and Dupont (1995)}
{\sc  Brenner, M. P.,  Lohse, D.,}
 and  {\sc Dupont, T. F.,}
``Bubble shape oscillations and the onset of sonoluminescence'',
  Phys. Rev. Lett. {\bf 75},  954-957
  (1995).

\bibitem{}{bre96}{Brenner et al. (1996a)}
{\sc Brenner, M.P., Lohse, D.,  Oxtoby, D.,}
 and  {\sc Dupont, T.F.,}
``Mechanisms for stable single bubble sonoluminescence'',
  Phys. Rev. Lett. {\bf 76},
  1158-1161  (1996).

\bibitem{}{bre96d}{Brenner et al. (1996b)}
{\sc  Brenner, M.~P.,  Rosales, R. R.,  Hilgenfeldt, S.,} and  
{\sc Lohse, D.,} 
``Acoustic energy storage in single bubble sonoluminescence'', preprint,
submitted to Phys. Rev. Lett., April 1996.

\bibitem{}{cru80}{Crum (1980)}
{\sc  Crum, L.~A.,}
``Measurements of growth of air bubbles by rectified diffusion'',
J. Acoust. Soc. Am. {\bf 68},  203-211  (1980).

\bibitem{}{cru94b}{Crum and Cordry (1994)}
{\sc  Crum, L.~A., } and 
{\sc Cordry, S.,}
``Single bubble sonoluminescence'',
  in {\em Bubble dynamics and interface phenomena},
  edited by J.~Blake et~al (Kluwer Academic Publishers, Dordrecht, 1994), p.\
  287-297.

\bibitem{}{ell69}{Eller (1969)}
{\sc Eller, A.,}
``Growth of bubbles by rectified diffusion'',
J. Acoust. Soc. Am. {\bf 46},  1246-1250  (1969).

\bibitem{}{ell70}{Eller and Crum (1970)}
{\sc Eller, A.,} and  {\sc Crum, L.A.,}
``Instability of the motion of a pulsating bubble in a sound field'',
 J. Acoust. Soc. Am. Suppl. {\bf 47},  762-767  (1970).

\bibitem{}{eps50}{Epstein and Plesset (1950)}
{\sc Epstein, P.S.,}  and  {\sc 
Plesset, M.S.,}
``On the stability of gas bubbles in liquid-gas solutions'',
J. Chem. Phys. {\bf 18},  1505-1509 (1950).

\bibitem{}{fre34}{Frenzel and Schultes (1934)}
{\sc Frenzel, H.,} and {\sc Schultes, H.,}
 Z. Phys. Chem. {\bf 27B},  421  (1934).

\bibitem{}{fyr94}{Fyrillas and Szeri (1994)}
{\sc Fyrillas, M.~M.,} and  {\sc Szeri, A.~J.,}
``Dissolution or growth of soluble spherical oscillating bubbles'',
 J. Fluid Mech. {\bf 277},  381-407  (1994).

\bibitem{}{gai90}{Gaitan et al. (1990)}
{\sc Gaitan, D.~F.,}
``An experimental investigation of acoustic cacitation in
               gaseous liquids'',
 Ph.D. thesis, The University of Mississippi, 1990;
{\sc  Gaitan, D.~F.,  Crum, L.~A.,  Roy,  R.~A.,} and  {\sc 
Church, C.~C.,}
Sonoluminescence and bubble dynamics for a single bubble, stable cavitation
bubble'',
 J. Acoust. Soc. Am. {\bf
  91},  3166-3183  (1992).

\bibitem{}{gre93}{Greenspan and Nadim (1993)}
{\sc  Greenspan, H.~P.,} and 
{\sc Nadim, A.,}
``On sonoluminescence of an oscillating gas bubble'',
Phys. Fluids A {\bf 5},  1065-1067  (1993).

\bibitem{}{bre96c}{Grossmann et al. (1996)}
{\sc  Grossmann, S.,  Hilgenfeldt, S.,  Lohse, D.,}
 and  {\sc Brenner, M.~P.,}
``Analysis of the Rayleigh-Plesset bubble dynamics for large forcing
pressure'', in preparation, May 1996.

\bibitem{}{hil96}{Hilgenfeldt, Lohse, and Brenner (1996)}
{\sc  Hilgenfeldt, S., Lohse, D.,} and  {\sc Brenner, M.~P.,}
``Phase diagrams for sonoluminescing bubbles'',
preprint, Phys.\ Fluids, November 1996.

\bibitem{}{hil92}{Hiller, Putterman, and Barber (1992)}
{\sc  Hiller, R.,  Putterman, S.~J.,}
 and  {\sc Barber, B.~P., }
 ``Spectrum of synchronous picosecond sonoluminescence'',
Phys. Rev. Lett. {\bf 69},  1182-1185
  (1992).

\bibitem{}{hil94}{Hiller et al. (1994)}
{\sc Hiller, R.,  Weninger, K.,  Putterman,  S.~J., }
and {\sc Barber, B.~P.,}
``Effect of noble gas doping in single bubble sonoluminescence'',
Science {\bf 266},
  248-234  (1994).

\bibitem{}{hil95}{Hiller and Putterman (1995)}
{\sc  Hiller, R.,}  and {\sc Putterman, S.~J.,}
``Observation of isotope effects in sonoluminescence''
 Phys. Rev. Lett. {\bf 75},  3549-3551  (1995).

\bibitem{}{lau76}{Lauterborn (1976)}
{\sc  Lauterborn, W.,}
``Numerical investigation of nonlinear oscillations of gas bubbles
in liquid'',
J. Acoust. Soc. Am. {\bf 59},  283-293  (1976).

\bibitem{}{loh96}{Lohse et al. (1996)}
{\sc Lohse, D.,
 Brenner, M.P., Dupont, T.,  Hilgenfeldt, S.,}
and  {\sc Johnston, B.,}
``Sonoluminescence: Air bubbles as chemical reaction chambers'',
preprint, April 1996.

\bibitem{}{loe93}{L\"ofstedt, Barber, and Putterman (1993)}
{\sc  L\"ofstedt, R.,  Barber, B.~P.,}
 and  {\sc Putterman, S.~J.,}
 ``Toward a hydrodynamic theory of sonoluminescence'',
Phys. Fluids A {\bf 5},  2911-2928
   (1993).

\bibitem{}{loe95}{L\"ofstedt et al. (1995)}
{\sc  L\"ofstedt, R.,  Weninger, K.,  Putterman,  S.~J.,}  and  
{\sc Barber,
``Sonoluminescing bubbles and mass diffusion''
B.~P.,} Phys. Rev. E
  {\bf 51},  4400-4410  (1995).

\bibitem{}{mos94}{Moss et al. (1994)}
{\sc Moss, W.,  Clarke, D.,  White, J., } and  {\sc Young, D.,}
``Hydrodynamic simulations of bubble collapse and picosecond
sonoluminescence'',
Phys. Fluids {\bf 6},  2979-2985 (1994).

\bibitem{}{ple49}{Plesset (1949)}
{\sc  Plesset, M.,}
``The dynamics of cavitation bubbles'',
J. Appl. Mech. {\bf 16},  277  (1949).

\bibitem{}{ple54}{Plesset (1954)}
{\sc  Plesset, M.,}
``On the stability of fluid flows with spherical symmetry'',
J. Appl. Phys. {\bf 25},  96-98  (1954).

\bibitem{}{ple77}{Plesset and Prosperetti (1977)}
{\sc  Plesset, M.} and  {\sc Prosperetti, A., }
``Bubble dynamics and cavitation''
 Ann. Rev. Fluid Mech. {\bf 9},  145-185  (1977).

\bibitem{}{pro77}{Prosperetti (1977)}
{\sc  Prosperetti, A.,}
``Viscous effects on perturbed spherical flows'',
 Quart. Appl. Math. {\bf 34},  339-352  (1977).

\bibitem{}{ray17}{Lord Rayleigh (1917)}
{\sc Rayleigh, Lord,}
``On the pressure developed in a liquid on the collapse
of a spherical bubble'',
 Philos. Mag. {\bf 34},  94  (1917).

\bibitem{}{str71}{Strube (1971)}
{\sc Strube, H.~W.,}
``Numerische Untersuchungen zur Stabilit\"at nichtsph\"arisch
schwingender Blasen'',
 Acustica {\bf 25},  289-303  (1971).

\bibitem{}{wen95}{Weninger et al. (1995)}
{\sc  Weninger, K.,  Hiller, R.  Barber, B.,  Lacoste, D.,}
  and {\sc Putterman, S.,}
``Sonoluminescence from single bubbles in non-aqueous liquids:
new parameter space for sonochemistry'',
 J. Phys. Chem. {\bf 99},  14195  (1995).

\bibitem{}{wu93}{Wu and Roberts (1993)}
{\sc  Wu, C.~C.,}  and  {\sc Roberts, P.~H.,}
``Shock-wave propagation in a sonoluminescing gas bubble'',
 Phys. Rev. Lett. {\bf 70},  3424-3427  (1993);
``A model of sonoluminescence'',
Proc. R. Soc. London A {\bf 445},  323-349  (1994).


\end{thebibliography}
\end{document}